\documentclass[12pt]{article}
\newcommand{\dd}{\mbox{\rm d}}

\newcommand{\oo}{\over}
\newcommand{\p}{\partial}

\newcommand{\be}{\begin{equation}}
\newcommand{\ee}{\end{equation}}
\newcommand{\lbl}{\label}
\newcommand{\bi}{\bibitem}
\newcommand{\ci}{\cite}

\newcommand{\vs}{\vspace}

\def\bear{\begin{eqnarray}}
\def\ear{\end{eqnarray}}

\begin{document}
\baselineskip .50cm

\thispagestyle{empty}

\ 

\vs{1.5cm}

\begin{center}

{\bf \Large CLIFFORD ALGEBRA, GEOMETRY AND PHYSICS}\footnote{Talk presented
at NATO Advanced Research Workshop ``The Nature of Time: Geometry,
Physics \& Perception", May 2002, Tatranska\' a Lomnica, Slovakia}

\vs{6mm}

Matej Pav\v si\v c\footnote{Email: MATEJ.PAVSIC@IJS.SI}

Jo\v zef Stefan Institute, Jamova 39, SI-1000 Ljubljana, Slovenia

\vs{6mm}

ABSTRACT

\end{center}

\vs{1mm}

The geometric calculus based on Clifford algebra is a very useful
tool for geometry and physics. It describes a geometric structure
which is much richer than the ordinary geometry of spacetime.
A Clifford manifold (C-space) consists not only of points, but also
of 1-loops, 2-loops, etc.. They are associated with multivectors which
are the wedge product of the basis vectors, the generators of Clifford
algebra. Within C-space we can perform the so called polydimensional
rotations which reshuffle the multivectors, e.g., a bivector into
a vector, etc.. A consequence of such a polydimensional rotation is
that the signature can change: it is relative to a chosen set of basis
vectors. Another important consequence is that the well known
unconstrained Stueckelberg theory is embedded within the constrained
theory based on C-space. The essence of the Stueckelberg theory is
the existence of an evolution parameter which is invariant under the
Lorentz transformations. The latter parameter is interpreted as being
the true time - associated with our perception of the passage of time.

\vs{4mm}

\section{Introduction}

In the usual theory of relativity there is no evolution. Worldlines
are fixed, everything is frozen once for all in a 4-dimensional
`` block universe" $V_4$. This is in contradiction with our 
subjective experience of the passage of time. It is in contradiction
with what we actually observe.

Therefore we always introduce into the theory of relativity more or less
explicitly {\it an extra postulate}: that a 3-dimensional hypersurface
of simultaneity {\it moves} in spacetime. We are talking about point
particles, strings, etc.. Those objects exist in 3-dimensional space
$V_3$. From the point of view of $V_4$ there are worldlines, worldsheets,
etc.. Relativity does not contain point particles that {\it evolve}
in $V_4$. Something is missing in the ordinary relativity. And yet,
we all: (i) assume the validity  of the theory of relativity, and
(ii) talk about point particles which ---when moving--- describe
worldlines in spacetime.

The above two positions are incompatible. In the following I am going to
point out how we can have both, (i) and (ii), by suitably modifying the
theory of relativity. The first modification is the well known Stueckelberg
theory \ci{Stueckelberg}, based on the unconstrained, Lorentz 
invariant action. Such a theory has been considered by a number of authors 
\ci{Feyman}--\ci{Pavsic1} and it actually
describes {\it evolution} of a point particle (``event") in spacetime.
The essence of the Stueckelberg theory is the introduction of a {\it
Lorentz invariant parameter} $\tau$ along which evolution (``relativistic
dynamics") takes place. This is the true {\it time}, whilst $X^0 \equiv t$ 
is just one of the spacetime coordinates, called ``coordinate time".

In search of a deeper understanding and description of geometry it has
been found that Clifford algebra is such a tool. It describes
a geometric structure which is much richer than the ordinary geometry
of spacetime. Clifford space (shortly $C$-space) consists not only of
points, but also of lines or 1-loops, 2-loops, etc.. The later geometric
objects are associated with multivectors. Multivectors of different
grades can be superposed into the geometric objects ---the so called
{\it polyvectors} which are generic {\it Clifford numbers} (called
also {\it Clifford aggregates}). Following Pezzaglia we assume that
physical quantities  are polyvectors and that the true space in which
physics stakes place is $C$-space \ci{Pezzaglia}--\ci{Pavsic3}.

We formulate the action in $C$-space, which is a straightforward although
not trivial generalization of the minimal length (point particle) action
in ordinary spacetime. Such {\it constrained} action contains 
as a particular case the well known
{\it unconstrained} Stueckelberg action which encompasses an invariant
{\it evolution parameter}. From the point of view of $C$-space, the above
evolution parameter is given by 4-vector part of the polyvector
describing position of the ``particle"\footnote{From the point of view
of spacetime, of course, this is not {\it particle}, but an aggregate
of $r$-loops, that is a polydimensional extended object 
(see \ci{Pezzaglia}--\ci{Pavsic3}).}
in $C$-space.

The theory of relativity is thus shifted from the ordinary spacetime
into the $C$-space. Everything that we know about relativity
is now true in $C$-space: the constrained minimal length action, invariance
under rotations (Lorentz transformations) in $C$-space, ``block universe",
etc.. But in spacetime, a subspace of $C$-space, particles (and also 
extended objects) are actually {\it moving} as suggested by the
unconstrained Stueckelberg action---which is just a reduced $C$-space
action.

\section{Relativistic Point Particle and Evolution}

We will now briefly review the Stueckelberg theory. Let us start
from the following action:
\be
    I = {1\oo 2} \int \dd \tau \left ( {{{\dot x}^\mu {\dot x}_\mu} \oo
    \Lambda} + \Lambda \kappa^2 \right )
\lbl{1}
\ee
where $\kappa$ is a constant. Let us consider two distinct procedures:

a) In the {\it standard procedure} $\Lambda$ is taken as a Lagrange
multiplier whose ``equations of motion" give $\Lambda^2 = {\dot x}^\mu
{\dot x}_\mu/\kappa^2$ which is equivalent to the {\it constraint}
$p^\mu p_\mu - \kappa^2 = 0$, where $p_\mu = \p L/\p {\dot x}^\mu =
{\dot x}_\mu/\Lambda$ is the canonical momentum. The action (\ref{1})
is then equivalent to the minimal length action $I = m \int \dd \tau
({\dot x}^\mu {\dot x}_\mu )^{1/2}$. Fixing $\Lambda$ in (\ref{1}) means
fixing a ``gauge", i.e., a choice of parametrization.

b) In the {\it non standard procedure} $\Lambda$ in (\ref{1}) is
taken to be a {\it constant} with a physical meaning. Here $\Lambda$ has
nothing to do with choice of parametrization. Then (\ref{1}) is an
{\it unconstrained} action and all $x^\mu$ are independent variables.
They satisfy the following equations of motion:
$(\dd/\dd \tau) {\dot x}^\mu/ \Lambda = 0$
where all $p_\mu = {\dot x}^\mu/ \Lambda$ are constants of motion, 
and so it is the square $p_\mu p^\mu = M^2$.

A particle's trajectory is given by $x^\mu(\tau)$. Here $x^0$ is one of the
coordinates, called {\it coordinate time}\footnote{
It is called ``clock time" by Franck \ci{Franck}.
},
whilst $\tau$ is the {\it evolution parameter} or {\it historical time}.
The variables $x^i (\tau), ~i=1,2,3$ describe the usual
spatial motion of the particle, ${\dot x}^i (\tau) \equiv \dd x^i/\dd \tau$
being the spatial velocity. The variable $x^0 (\tau)$ describes the
progression of particle's coordinate $x^0$ with increasing evolution
parameter $\tau$; ${\dot x}^0 (\tau)$ is the {\it speed} of the coordinate
time with respect to the evolution parameter $\tau$. The latter parameter we
interpret as being related to the time perceived by consciousness
when experiencing the passage of time. A given value of $\tau$ denotes
{\it ``now"}, whilst $x^0 (\tau)$ (together with $x^i (\tau)$) denotes 
{\it position} in spacetime. {\it The Stueckelberg theory as interpreted
in} \ci{Pavsic1,Pavsic2} {\it thus describes progression of ``now",
the concept which is not present in the ordinary theory of relativity.}
This is even more transparent in the quantized theory.

From (\ref{1}) it is straightforward to derive the Hamiltonian
\be
     H = p_\mu {\dot x}^\mu - L = {\Lambda\oo 2} (p^\mu p_\mu - \kappa^2) .
\lbl{4}
\ee
Since $\kappa$ is an arbitrary constant, it can be taken $\kappa=0$ 
(as it is in the usual formulation of the Stueckelberg theory).

In the {\it quantized theory} $x^\mu, ~p_\mu$ become operators, satisfying
$[x^\mu, p_\mu] = i {\delta^\mu}_\nu$  ($\hbar = c = 1$).
In the representation in which $x^\mu$ are diagonal, momenta are
$p_\mu = - i \p_\mu$. A state can be represented by a wave function
$\psi(\tau, x^\mu)$ satisfying the Schr\" odinger equation
$i\p \psi/\p \tau = H \psi$.

The wave function is normalized in {\it spacetime} according to
$\int \psi^* \psi \, \dd^4 x = 1$.
The latter relation holds at any value of $\tau$. Therefore the evolution
operator $U$ which sends $\psi (\tau)$ into $\psi (\tau') = U \psi (\tau)$
is {\it unitary}. In other words, because of the above normalization
{\it unitarity is satisfied even if wave functions are localized
in the coordinate time $x^0$}.

A generic wave function---a wave packet localized in spacetime---is a
superposition of the wave functions with definite 4-momentum:
\be
     \psi(\tau,x) = \int \dd^4 p \, c(p) \, {\rm exp} \left [ i p_\mu
     x^\mu - i {\Lambda \oo 2} (p^2 - \kappa^2) \tau \right ] .
\lbl{8}
\ee
The function $c(p)$ determines the profile of the wave packet.
Here both $p_\mu$ and its square $p^\mu p_\mu = M^2$ are indefinite.
  
In general, a state $\psi (\tau,x)$ has {\it indefinite} mass; the wave
packet is localized in spacetime.
The region of localization depends on the evolution parameter $\tau$.
The centre of the wave packet describes a classical world line (see
figures in \ci{Pavsic1,Pavsic3}).

It is now natural to interpret the wave function localized in spacetime
as being related to our perception of ``now" \ci{Pavsic1,Pavsic3}.
When the wave packet evolves with $\tau$, its region of localization
(center of the wave packet) {\it moves} in spacetime along a time-like
direction. This is then a physical description of the ``passage of time".

At this point let me mention that the wave function for a {\it localized point
particle} (``event") in spacetime is just a first step. Instead of localized
point particles we can consider {\it localized
extended objects} (strings, membranes) in spacetime whose dynamics is
given in terms of wave functionals satisfying the unconstrained
Schr\" odinger functional equation \ci{Pavsic4}.

An example is a {\it string} extended along a {\it time-like} direction.
Such time-like strings, if charged, yield the correct electromagnetic 
interaction with the Coulomb law.\footnote{Charged point particles localized 
in spacetime (charged ``events") do not lead to the Coulomb law.}

Another example is a 4-dimensional membrane ${\cal V}_4$ in an $N$-dimensional
embedding space. According the ``brane world" scenario such a
membrane ${\cal V}_4$ could be our world\footnote{In \ci{Pavsic6}
it was shown that self-intersections of ${\cal V}_4$ (or the intersections
of ${\cal V}_4$ with other branes) give rise to {\it localized
matter} on ${\cal V}_4$.}. Quantum mechanically motion of  ${\cal V}_4$
is described by a wave functional which can be sharply localized within a 
certain 4-region $\Omega$ on ${\cal V}_4$. Such region could correspond
to ``here" and ``now". With the passage of $\tau$ the wave functional
evolves so that the region of sharp localization $\Omega$ changes and so
also ``here" and ``now" change. In short, I assume the interpretation
that such localized wave functional provides a physical description of
our perception of ``here" and ``now". Much more on this topics is to be
found in a recent book \ci{Pavsic3} and in \ci{Pavsic5}.

\section{Geometric Calculus Based on Clifford Algebra}

I am going to provide a brief, simplified, introduction into the calculus
with {\it vectors} and their generalizations.\footnote{A more elaborate
discussion is in \ci{Pavsic3}.} Geometrically, a vector is an
oriented line element. Mathematically, it can be elegantly described
as a {\it Clifford number} \ci{Hestenes}.

How to multiply vectors? There are two possibilities:

1. {\it The inner product}
\be
      a \cdot b = b \cdot a
\lbl{3.1}
\ee
of vectors $a$ and $b$. The quantity $a \cdot b$ is a {\it scalar}.

2. {\it The outer product}
\be
      a \wedge b = - b \wedge a
\lbl{3.2}
\ee
which is an oriented element of a plane.

 The products 1 and 2 can be considered as the {\it symmetric} and the
 {\it anti symmetric} parts of the {\it Clifford product}, called also
 {\it geometric product}
 \be
        a b = a \cdot b + a \wedge b
\lbl{3.3}
\ee
where
\be
     a \cdot b \equiv \mbox{${1\oo 2}$} (a b + b a), \qquad
       a \wedge b \equiv \mbox{${1\oo 2}$} (a b - b a).                        
\lbl{3.5}
\ee

This suggests a generalization to trivectors, quadrivectors, etc.. It
is convenient to introduce the name $r$-{\it vector} and call $r$ its
{\it degree}: $A_r = a_1 \wedge a_2 \wedge ... \wedge a_r$. Another name
for a generic $r$-vector is {\it multivector}. The highest possible
multivector in $V_n$ is $n$-vector, since $(n+1)$-vector is identically zero.

Let $e_1, \, e_2, \, ..., \, e_n$ be linearly independent vectors,
and $\alpha, \, \alpha^i, \, \alpha^{i_1 i_2}, ...$ scalar coefficients.
A generic Clifford number can then be written as
\be
       A = \alpha + \alpha^\mu e_\mu + \mbox{${1\oo {2!}}$}
        \, \alpha^{\mu_1 \mu_2}
       \,e_{\mu_1} \wedge e_{\mu_2} + ... \mbox{${1\oo {n!}}$}
        \, \alpha^{\mu_1 ... \mu_n}
       e_{\mu_1} \wedge ... \wedge e_{\mu_n} .
\lbl{3.6}
\ee

Since it is a superposition of multivectors of all possible grades
it will be called {\it polyvector}.\footnote{
Following a suggestion by Pezzaglia \ci{Pezzaglia} I call a generic 
Clifford number
{\it polyvector} and reserve the name {\it multivector} for an $r$-vector,
since the latter name is already widely used for the corresponding object
in the calculus of differential forms.} Another name, also often used
in the literature, is {\it Clifford aggregate}. These mathematical objects
have far reaching geometrical and physical implications that will be
discussed and explored to some extent in the rest of the paper.

In general $e_\mu$ in eq.(\ref{3.6}) are arbitrary so that their inner
products form the metric tensor of arbitrary signature. In particular $e_\mu$
can be four linearly independent vectors $e_\mu = \gamma_\mu , ~
\mu = 0,1,2,3$, satisfying $\gamma_\mu \cdot \gamma_\nu = \eta_{\mu \nu}$,
generating the Clifford algebra of spacetime, called the {\it Dirac algebra}.
By using the relations $\gamma_{\mu \nu \rho \sigma} = \gamma_5
\epsilon_{\mu \nu \rho \sigma}$ and $\gamma_{\mu \nu \rho} = \gamma_{\mu \nu
\rho \sigma} \gamma^\sigma$, where $\gamma_{\mu \nu ...} \equiv
\gamma_\mu \wedge \gamma_\nu \wedge ...$ we can rewrite $A$ as a superposition
of a scalar, vector, bivector, pseudovector and pseudoscalar:
\be
    A = S + V^{\mu} \gamma_{\mu} + T^{\mu \nu} \gamma_{\mu \nu} + 
    C^{\mu} \gamma_5 \gamma_{\mu} + P \gamma_5
\lbl{3.7}
\ee
where $ S \equiv \alpha ,~ V^{\mu} \equiv \alpha^{\mu}, 
~T^{\mu \nu}
\equiv (1/2) \alpha^{\mu \nu},~C_{\sigma} \equiv 
(1/3!) \alpha^{\mu \nu \rho} 
    \epsilon_{\mu \nu \rho \sigma}$ and $P \equiv (1/4!)
    \alpha^{\mu \nu \rho \sigma} \epsilon_{\mu \nu \rho \sigma}$.
    
\vs{2mm}    
    
{\it Relativity of signature.}  In eq.(\ref{3.7}) we assumed the Minkowski
signature of the metric tensor. We are now going to find out that within
Clifford algebra the signature is a matter of which amongst the available
Clifford numbers we choose as the {\it basis vectors} (i.e., as the
generators of Clifford algebra).

Let us assume that the basis vectors $e_\mu , ~ \mu = 0,1,2,3$, satisfy
\be
    e_\mu \cdot e_\nu \equiv \mbox{${1\oo 2}$} (e_\mu e_\nu + e_\nu e_\mu) =
    \delta_{\mu \nu}
\lbl{3.8}
\ee
where $\delta_{\mu \nu}$ is the {\it Euclidean metric}.

Let us consider the set of four Clifford numbers $(e_0,\, e_i e_0)$,
$i = 1,2,3$ and denote them as
\be 
      e_0 \equiv \gamma_0 \; , \qquad e_i e_0 \equiv \gamma_i .
\lbl{3.9}
\ee 
The Clifford numbers $\gamma_{\mu}$, $\mu = 0,1,2,3$ satisfy
\be
     \gamma_\mu \cdot \gamma_\nu =
         \mbox{${1\oo 2}$ }(\gamma_{\mu} \gamma_{\nu} + 
         \gamma_{\nu} \gamma_{\mu})
         = \eta_{\mu \nu}
\lbl{3.10}
\ee
where $\eta_{\mu \nu} = {\rm diag} (1, -1, -1 ,-1)$ is the {\it Minkowski
tensor}. We see that the $\gamma_{\mu}$ behave as basis vectors in a
4-dimensional space $V_{1,3}$ with signature $(+ - - -)$. We can form a
Clifford aggregate $\alpha = \alpha^{\mu} \gamma_{\mu}$
which has the properties of a {\it vector} in $V_{1,3}$. From the point of
view of the space $V_4$ the same object $\alpha$ is a linear combination of
a vector and bivector:
$\alpha = \alpha^0 e_0 + \alpha^i e_i e_0$.

We may use $\gamma_{\mu}$ as generators of the Clifford algebra
${\cal C}_{1,3}$ defined over the pseudo-Euclidean space $V_{1,3}$. The
basis elements of ${\cal C}_{1,3}$ are $\gamma_J =$\ $(1,\gamma_{\mu},\gamma_{\mu
\nu}, \gamma_{\mu \nu \alpha}, \gamma_{\mu \nu \alpha \beta})$, with
$\mu < \nu < \alpha < \beta$. A generic  Clifford aggregate in ${\cal C}_{1,3}$
is given by
\be
        B = b^J \gamma_J = b + b^{\mu} \gamma_{\mu} + 
        b^{\mu \nu} \gamma_{\mu} \gamma_{\nu} +
        b^{\mu \nu \alpha} \gamma_{\mu} \gamma{\nu} \gamma_{\alpha} + 
        b^{\mu \nu \alpha \beta}
        \gamma_{\mu} \gamma_{\nu} \gamma_{\alpha} \gamma_{\beta} .
\lbl{3.12}
\ee
With suitable choice of the coefficients $b^J = (b, b^{\mu},
b^{\mu \nu}, b^{\mu \nu \alpha}, b^{\mu \nu \alpha \beta})$
we have that $B$ of eq.(\ref{3.12}) is equal to $A$ of eq.(\ref{3.6}). Thus the
same number $A$ can be described either within ${\cal C}_4$ or within
${\cal C}_{1,3}$. The expansions (\ref{3.12}) and (\ref{3.6}) 
exhaust all possible
numbers of the Clifford algebras ${\cal C}_{1,3}$ and ${\cal C}_4$.
The algebra ${\cal C}_{1,3}$ is isomorphic to the
algebra ${\cal C}_4$ and actually they are just two different representations
of the same set of Clifford numbers (called also polyvectors or Clifford
aggregates).

\section{Extending Relativity from Spacetime to $C$-space}

So far it has been assumed that the arena in which physics takes place is
spacetime. The nice properties of Clifford algebra suggest to extend the
arena to a larger manifold, called {\it Clifford space} or $C$-{\it space},
whose points are described by {\it coordinate polyvectors}
\be
      X = {1\oo {r!}} \sum_{r=0}^n X^{\mu_1...\mu_r} \gamma_{\mu_1} \wedge
      ...\wedge \gamma_{\mu_r} \equiv X^A E_A .
\lbl{4.1}
\ee
Here $X^A$ are coordinates, and $E_A = ({\underline 1}, \gamma_\mu, 
\gamma_\mu \wedge \gamma_\nu,...)$ basis vectors of $C$-space.

Points $X$ of $C$-space embrace, from the spacetime perspective, not only
the usual points, but also 1-loops, 2-loops, etc.. Thus $C$-space is a
polydimensional continuum \ci{Pezzaglia} in which the extended object 
of different dimensionalities coexist on the same footing, and can be
transformed into each other by {\it polydimensional rotations} which are
a generalization of Lorentz transformations\footnote{Here I am
considering flat $C$-space. Curved $C$-space is considered in 
\ci{Pavsic3}, \ci{Castro-Pavsic}.}.

The line element in $C$-space is given by the scalar product of an
infinitesimal polyvector $\dd X = \dd X^A E_A$ and its reverse 
$\dd X^{\dagger}$:
\be
       |\dd X |^2 \equiv \dd X^{\dagger} * \dd X = \dd X^A \, \dd X^B
       G_{AB} = \dd X^A \, \dd X_A
\lbl{4.2}
\ee
where $G_{AB} = E_A^{\dagger} * E_B$
is the {\it metric} of $C$-space. The {\it scalar product} of two
polyvectors $A$ and $B$ is defined as the scalar part of the Clifford
product $AB$, i.e., $A * B = \langle AB \rangle_0$. The symbol $A^{\dagger}$
denotes the {\it reverse} of $A$, that is the polyvector in which
the order of all products of vectors in a decomposition of a polyvector
$A$ is reverse (e.g., $(\gamma_1 \gamma^2 \gamma^3)^\dagger = 
\gamma_1 \gamma^2 \gamma^3$).

The reparametrization invariant action for a point particle in $C$-space is
\be
       I[X^A] = \kappa \int \dd \tau \, ({\dot X}^A {\dot X}_A )^{1/2}
\lbl{4.4}
\ee
where $\kappa$ is a constant. Here ${\dot X}^A \equiv \dd X^A/\dd \tau$,
where $\tau$ is an arbitrary parameter. The canonical momenta are
$p_A = \kappa {\dot x}_A/({\dot x}^B {\dot x}_B)^{1/2}$ and they
satisfy the constraint $p^A p_A = \kappa^2$.

Let us assume that spacetime dimension is $n=4$. Then the velocity polyvector
is
\be
    {\dot X} \equiv {\dot X}^A E_A = {\dot \sigma} {\underline 1} +
    {\dot x}^\mu \gamma_\mu + \mbox{${1\oo 2}$} {\dot x}^{\mu \nu} \gamma_\mu
    \wedge \gamma_\nu + {\dot \xi}^\mu \gamma_5 \gamma_\mu + {\dot s}
    \gamma_5 .
\lbl{4.5}
\ee
From the action (\ref{4.4}) it follows that ${\dot X}$ is constant when
$C$-space is flat.

In particular, when the initial conditions happen to be such that 
${\dot \sigma} =0, ~ {\dot x}^{\mu \nu} = 0 , ~ {\dot \xi}^{\mu} = 0$, we have
${\dot X} = {\dot x}^\mu {\dot s} \gamma_5$, $|{\dot X}|^2 \equiv
{\dot X}^A {\dot X}_A = {\dot x}^\mu {\dot x}_{\mu} - {\dot s}^2$, and the
action takes the form
\be
    I[x^\mu, s] = \kappa \int \dd \tau ({\dot x}^\mu {\dot x}_\mu -
    {\dot s})^{1/2} .
\lbl{4.6}
\ee
    
Not all the variables $X^A (\tau)$ in (\ref{4.4}) are independent dynamical
degrees of freedom. Here $x^\mu (\tau), ~ s(\tau)$ in (\ref{4.6}) also are
not all independent. But we can fix a gauge (choose a parametrization) and
thus reduce the set of variables to the physical, independent,
dynamical variables. A natural choice of gauge in (\ref{4.6}) is
$s = \tau$. The reduced action is then
\be
     I[x^\mu] = \kappa \int \dd s \, ({\dot x}^\mu {\dot x}_\mu - 1)^{1/2} .
\lbl{4.7}
\ee
The equations of motion derived from (\ref{4.7}) are
\be
    {{\dd p_\mu}\oo {\dd s}} = 0 , \qquad p_\mu = {{\kappa {\dot x}_\mu}
    \oo {({\dot x}^\nu {\dot x}_\nu - 1)^{1/2}}} = constant .
\lbl{4.8}
\ee
The square $p^\mu p_\mu = M^2 = \kappa^2  {\dot x}^\mu {\dot x}_\mu /
({\dot x}^\nu {\dot x}_\nu - 1)^{1/2}$ is also a constant of motion.
Inserting the latter relation between $M$ and $\kappa$ into the expression
(\ref{4.8}) we obtain
\be
    p_\mu = {{M {\dot x}_\mu}\oo {({\dot x}^\nu {\dot x}_\nu)^{1/2}}} ,
    \qquad M = \kappa {{({\dot x}^\mu {\dot x}_\mu)^{1/2}}\oo 
    {({\dot x}^\nu {\dot x}_\nu - 1)^{1/2}}} .   
\lbl{4.9}
\ee
The latter expression for 4-momentum has formally the same form as the momentum
of the usual 4-dimensional relativity. The difference is in that $M$ is not
a fixed constant entering the action, but a constant of motion. But the
latter property is just typical for the {\it Stueckelberg unconstrained
theory} described by the action (\ref{1}).

We shall now directly demonstrate that the constrained action (\ref{4.6})
is equivalent to the Stueckelberg action (\ref{1}). First we observe that
instead of (\ref{4.6}) we can use the Howe-Tucker type action in which
there occurs a Lagrange multiplier $\lambda$:
\be
      I[x^\mu, s,\lambda] = {\kappa \oo 2} \int \dd \tau \, 
      \left ( {{{\dot x}^\mu {\dot x}_\mu - {\dot s}^2}\oo \lambda} + \lambda
      \right )
\lbl{4.10}
\ee
which is classically equivalent to (\ref{4.6}). Variation of (\ref{4.10})
with respect to $x^\mu, ~s, ~\lambda$ gives
\be      
      {\dd \oo {\dd \tau}} \left ( {{\kappa {\dot x}^\mu}\oo \lambda }
      \right ) = 0 ,  \qquad
      {\dd \oo {\dd \tau}} \left ( {{\kappa {\dot s}}\oo \lambda }
      \right ) = 0  , \qquad \lambda = ({\dot x}^\mu {\dot x}_\mu -
      {\dot s}^2)^{1/2} .
\lbl{4.11}
\ee
The second equation in (\ref{4.11}) gives $(\dd/\dd \tau)(\kappa {\dot s} 
s/\lambda) = \kappa {\dot s}^2/\lambda$. Using the latter equation we can
rewrite eq.(\ref{4.10}) in the form
\be
     I[x^\mu,s,\lambda] = {\kappa\oo 2} \int \dd \tau \, \left [
     {{{\dot x}^\mu {\dot x}_\mu}\oo \lambda} + \lambda -
     {\dd \oo {\dd \tau}} \left ( {{\kappa {\dot s} s}\oo \lambda} \right )
     \right ] .
\lbl{4.12}
\ee
The Lagrange multiplier $\lambda$ can be chosen arbitrarily: this determines
a choice of parametrization. Let us choose $\lambda = \Lambda \kappa$,
i.e., $({\dot x}^\mu {\dot x}_\mu - {\dot s}^2)^{1/2} = \Lambda \kappa$,
where $\lambda$ is a fixed {\it constant}. Omitting the total derivative,
eq.(\ref{4.12}) becomes just
the {\it Stueckelberg action} (\ref{1})! The equations of motion derived 
from the unconstrained action (\ref{1}) are the same as the $x^\mu$
equations (\ref{4.11}) derived from the constrained action (\ref{4.10}).

\section{Conclusion}

The formulation of relativity in $C$-space leads to the point particle with
an extra variable $s$ along which the evolution in spacetime takes
place. The extra variable $s$ does not come from an extra dimension
of spacetime $V_4$, but from the Clifford algebra of $V_4$. In $C$-space
we have ``block universe", no evolution, everything frozen. But in
Minkowski space $V_4$ we have evolution. All the elegance of the
theory of relativity is preserved, not in $V_4$, but in $C$-space.
All the nice features of the Stueckelberg unconstrained theory are also
present, not in $C$-space, but in its subspace $V_4$.

It is often claimed that the passage of time is just an illusion of
the observer. Well, but good physics has always been capable of
explaining certain illusions. Physics did not raise hands at why we
see a ``lake" in a desert, a colored arc in the rainy and sunny sky,
or why far away objects appear smaller than the nearby ones. Now it is
time to explain why we experience the passage of time. In the present
paper I have presented a theoretical framework in which such a problem
could be tackled.

\end{document}